\documentclass[12pt,twoside]{article}
\usepackage{amssymb}
\newcommand{\be}{\begin{eqnarray}}
\newcommand{\ee}{\end{eqnarray}}
\newcommand{\pa}{\partial}
\newcommand{\A}{{\cal A}}
\newcommand{\B}{{\cal B}}
\newcommand{\la}{\lambda}
\newcommand{\htimes}{\hat{\times}}

\title{\bf Algebraic identities associated with \\ KP and AKNS hierarchies
       \thanks{\copyright 2005 by A. Dimakis and F. M\"uller-Hoissen} }
\author{Aristophanes Dimakis \\
 Department of Financial and Management Engineering, \\
 University of the Aegean, 31 Fostini Str., GR-82100 Chios, Greece \\
 dimakis@aegean.gr
          \and
 Folkert M\"uller-Hoissen \\ Max-Planck-Institute for Dynamics and Self-Organization \\
 Bunsenstrasse 10, D-37073 G\"ottingen, Germany \\
 fmuelle@gwdg.de }

\date{}

\begin{document}

\maketitle

\begin{abstract}
Explicit KP and AKNS hierarchy equations can be constructed from a certain set
of algebraic identities involving a quasi-shuffle product.
\end{abstract}

\section{Introduction}
The equations of the KP hierarchy are well-known to possess multi-soliton solutions.
According to Okhuma and Wadati \cite{Okhu+Wada83}, these solutions can be expressed
as formal power series (in some indeterminate). Substitution into hierarchy equations
leads to algebraic sum identities of a special kind. The structure of these identities
has been explored in \cite{DMH05} and abstracted to a certain algebra which we briefly
recall in section~\ref{section:algebra}.\footnote{These identities are in fact identities
of quasi-symmetric functions, see \cite{qsym} for example.}
Moreover, a map $\Phi$ from the latter algebra
to the algebra of pseudo-differential operators underlying the Gelfand-Dickey
formulation \cite{Dick03} of the KP hierarchy was constructed, which maps a certain
set of algebraic identities to KP hierarchy equations, and the whole KP hierarchy is
actually obtained in this way.
We briefly recall this map in section~\ref{section:KP}. In section~\ref{section:AKNS}
we show that, quite surprisingly, the same set of identities is also related to the AKNS
hierarchy in a similar way. In particular, relations between AKNS and KP emerge from
this relation, as will be demonstrated in section~\ref{section:AKNS->KP}.

\section{The algebra}
\label{section:algebra}
Let $\A = \bigoplus_{r \geq 1} \A^r$ be a graded linear space over
a field $\mathbb{K}$ of characteristic zero, supplied with two
products $\prec \, : \A^r \times \A^s \rightarrow \A^{r+s}$ and
$\bullet : \A^r \times \A^s \rightarrow \A^{r+s-1}$ which are
associative and also mixed associative. We assume\footnote{This assumption
should have been added in \cite{DMH05}.}
that $\A$ is generated by $\A^1$ via the product $\prec$.
Now we define a \emph{quasi-shuffle product}
(see, e.g., \cite{Guo+Keig00shuffle}) $\circ$ in $\A$ by
\be
    A \circ B &=& A \prec B + B \prec A + A \bullet B \\
    A \circ (B \prec \alpha) &=& A \prec B \prec \alpha + B \prec (A \circ \alpha)
      + A \bullet B \prec \alpha \\
   (A \prec \alpha) \circ B &=& A \prec (\alpha \circ B) + B \prec A \prec \alpha
      + A \bullet B \prec \alpha \\
     (A \prec \alpha) \circ (B \prec \beta)
 &=& A \prec ( \alpha \circ ( B \prec \beta) )
     + B \prec ( (A \prec \alpha) \circ \beta ) \nonumber \\
 & & + (A \bullet B) \prec ( \alpha \circ \beta )
\ee
for all $A,B \in \A^1$ and $\alpha, \beta \in \A$.
This is another associative product in $\A$ which, however, is \emph{not}
mixed associative with the other products. If $(\A^1, \bullet)$ is commutative,
then also $(\A, \circ)$ \cite{DMH05}.
\vskip.1cm

Let $\A(P)$ be the subalgebra generated by a single element $P \in \A^1$. We set
$P^{\bullet n} := P \bullet \ldots \bullet P$ ($n$-fold product).
Introducing the combined associative product $\succ := \prec + \, \bullet$, the
next formula defines a further associative product in $\A(P)$,
\be
     \alpha \htimes \beta := - \alpha \prec P \succ \beta  \; .
\ee
In the following $\cal I$ denotes the set of identities in $\A(P)$ built from the
elements $P^{\bullet n}$, $n=1,2, \ldots$ solely by use of the products
$\circ$ and $\htimes$.

\section{From $\A(P)$ to the KP hierarchy}
\label{section:KP}
Let $\mathcal{R}$ denote the $\mathbb{K}$-algebra of formal
pseudo-differential operators ($\Psi$DOs) generated by
\be
    L = \pa + \sum_{n \geq 1} u_{n+1} \, \pa^{-n}
\ee
with coefficients from an associative algebra $\B$ (over $\mathbb{K}$),
together with the projection $( \, )_{< 0}$ to that part of a $\Psi$DO
containing only negative powers of the partial derivative
operator $\pa$ (with respect to a variable $x$). Now
\be
    \ell(P) := L \, , \quad
    \ell(\alpha \prec P) := - \ell(\alpha)_{<0} \, L \, , \quad
    \ell(\alpha \bullet P) := \ell(\alpha) \, L
\ee
determines iteratively a map $\ell : \A(P) \rightarrow \mathcal{R}$. The map
$\Phi : \A(P) \rightarrow \B$ defined by
\be
    \Phi(\alpha) := \mbox{res}(\ell(\alpha))
\ee
(where the residue of a $\Psi$DO is the coefficient of its $\pa^{-1}$ term) then
has the following properties \cite{DMH05},
\be
    \Phi(\alpha \htimes \beta) = \Phi(\alpha) \, \Phi(\beta)  \, , \qquad\quad
    \Phi(P^{\bullet n} \circ \alpha) = \delta_n \Phi(\alpha)
\ee
where $\delta_n$ are derivations with $\delta_n L := - [ (L^n)_{<0} , L ]$. They should
be regarded as vector fields on the algebra $\mathcal{R}$.
 From \cite{DMH05} we recall:
\vskip.1cm
\noindent
{\bf Theorem.} Writing $u_2 = \phi_x$ and imposing the flow equations $\delta_n = \pa_{t_n}$
(where $t_1 = x$), which imply $\Phi(P^{\bullet n}) = \phi_{t_n}$,
all (combinations of) equations of the KP hierarchy lie in $\Phi({\cal I})$.
\hfill $\blacksquare$
\vskip.1cm

We believe that \emph{any} identity from $\cal I$ is mapped to a combination of KP
hierarchy equations, so that the correspondence is actually one-to-one.
This has not yet been proven, however. The set of algebraic identities specified in the
theorem expresses the `building rules' of \emph{explicit} KP hierarchy equations, which
are rather implicitly determined by the sequence of Lax equations
$L_{t_n} = - [ (L^n)_{<0} , L ]$ \cite{Dick03}. For example, the algebraic identity
\be
     4 \, P^{\bullet 3} \circ P - P^{\circ 4} - 6 \, P \circ ( P \htimes P )
   = 6 \, [ P^{\bullet 2} , P ]_{\htimes} + 3 \, P^{\bullet 2} \circ P^{\bullet 2}
     \label{KP-id}
\ee
is mapped by $\Phi$ to the (potential) KP equation
\be
  ( 4 \, \phi_{t_3} - \phi_{xxx} - 6 \, (\phi_x)^2 )_x
   =  6 \, [\phi_{t_2} , \phi_x ] + 3 \, \phi_{t_2t_2}  \; .    \label{pKP}
\ee

\section{From $\A(P)$ to the AKNS hierarchy}
\label{section:AKNS}
Let $\B$ be an associative $\mathbb{K}$-algebra with unit $I$,
and $\B_\la$ the algebra of formal series (in an indeterminate $\lambda$ and its inverse)
\be
    X = \sum_{m \leq M} \la^m X_m
\ee
where $X_m \in \B$ and $M \in \mathbb{Z}$. We set
\be
    X_{\geq 0} := \sum_{0 \leq m \leq M} \la^m X_m \, , \qquad
    X_{<0} := X-X_{\geq 0} = \sum_{m<0} \la^m X_m \; .
\ee
Next we choose an element $V \in \B_\la$ of the form
\be
    V = v_0 + \la^{-1} \, v_1 + \la^{-2} \, v_2 + \la^{-3} \, v_3 + \ldots \, , \qquad
    J := v_0
\ee
with $J,v_m \in \B$. Note that $v_m = V_{-m}$, $m=0,1,\ldots$.
A generalization of the well-known AKNS hierarchy (see also \cite{Dick03}) is then
determined by
\be
    V_{t_n} = [(\la^n V^n)_{\geq 0},V] = - [(\la^n V^n)_{<0},V] =: \delta_n V
    \qquad\quad   n = 1,2, \ldots             \label{gAKNS}
\ee
which requires $J_{t_n} = 0$. By a standard argument, the flows commute.
The first ($n=1$) hierarchy equation is equivalent to
\be
   [J,v_{m+1}] + [v_1,v_m] = v_{m,x} \qquad\quad m=1,2, \ldots \; .
   \label{gAKNS1}
\ee

Next we define two maps $\ell, \mathbf{r} : \A(P) \rightarrow \B_\lambda$ via
$\ell(P) = \mathbf{r}(P) = \la \, V$ and
\be
    \ell(\alpha\prec P) = -\ell(\alpha)_{<0} \, \la V \, , \quad && \quad
    \ell(\alpha \bullet P) = \ell(\alpha) \, \la V  \\
    \mathbf{r}(P \prec \alpha) = -\la V \, \mathbf{r}(\alpha)_{\geq 0} \, , \quad && \quad
    \mathbf{r}(P \bullet \alpha) = \la V \, \mathbf{r}(\alpha)
\ee
for all $\alpha \in \A(P)$. As a consequence, we have
$\ell(P^{\bullet m}) = \mathbf{r}(P^{\bullet m}) = \la^m \, V^m$.\footnote{Several properties
of $\ell$ and $\mathbf{r}$ in the KP case have been derived in \cite{DMH05}. Most of them
in fact remain valid if we replace $L$ by $\lambda \, V$.}
The map $\Phi : \A(P) \rightarrow \B$ defined by
\be
    \Phi(\alpha) := (\ell(\alpha)_{<0} \, \la V)_{\geq 0}
                  = \ell(\alpha)_{-1} \, J
\ee
then satisfies $\Phi(\alpha) = (\mathbf{r}(\alpha)_{<0} \, \la V)_{\geq 0}
= \mathbf{r}(\alpha)_{-1} \, J$ and
\be
    \Phi(P^{\bullet k})
  = (\la^k V^k)_{-1} \, J
  = (V^k)_{-(k+1)} \, J  \; .
\ee
In particular, we obtain $\Phi(P) = v_2 \, J$ and
\be
 \Phi(P^{\bullet 2}) &=& ( \{J,v_3\} + \{v_1,v_2\} ) \, J  \label{Phi(P2)} \\
 \Phi(P^{\bullet 3}) &=& ( \frac{1}{2} \{J,J,v_4\} + \{J,v_1,v_3\} + \frac{1}{2} \{J,v_2,v_2\}
   + \frac{1}{2} \{v_1,v_1,v_2\} ) \, J       \label{Phi(P3)}
\ee
where $\{ a_1, \ldots , a_k \} := \sum_{\sigma \in \mathcal{S}_k} a_{\sigma(1)}
 \cdots a_{\sigma(k)}$ with the symmetric group ${\cal S}_k$ of order $k$.
It can be shown that $\Phi$ has the following algebra homomorphism property,
\be
    \Phi(\alpha \htimes \beta)
  = (\ell(\alpha)_{<0} \, \la V \, \mathbf{r}(\beta)_{<0})_{-1} \, J
  = \Phi(\alpha) \, \Phi(\beta)
\ee
for all $\alpha, \beta \in \A(P)$ (cf theorem~6.2 in \cite{DMH05}).
Another important formula is\footnote{This is the analog of the simplest case expressed
by proposition~6.5 in \cite{DMH05}.}
\be
   \Phi(\alpha)_{t_n} = \delta_n \Phi(\alpha) = \Phi(P^{\bullet n} \circ \alpha)
\ee
where we imposed the flow equations $\delta_n = \pa_{t_n}$ on $\B_\la$.
Applying $\Phi$ to the simple algebraic identities
$P^{\bullet k} \circ P^{\bullet n} = P^{\bullet n} \circ P^{\bullet k}$ leads
to the relations
\be
   ( \Phi(P^{\bullet n}) )_{t_k} = ( \Phi(P^{\bullet k}) )_{t_n}  \label{Phi(PkPn)}
\ee
(which in the KP case are trivially satisfied). The identity (\ref{KP-id}) is
mapped by $\Phi$ to
\be
 && \Big( 4 \, v_{2,t_3} - v_{2,xxx} - 3 \, ( \{J,v_3\} + \{v_1,v_2\} )_{t_2}
  - 6 \, (v_2 J v_2)_x \Big) \, J  \nonumber  \\
 && + 6 \, [ v_2 J , (\{J,v_3\} + \{v_1,v_2\}) \, J ] = 0 \; .
  \label{KP-id-gAKNS}
\ee

\subsection{$V^2 = V$ reduction}
For any polynomial $\cal P$ of $V$ with coefficients in the center of $\B$,
the constraint ${\cal P}(V) = 0$, which in particular requires ${\cal P}(J) = 0$,
is preserved by the hierarchy (\ref{gAKNS}). Let us consider the special case
$V^2 = V$, which is equivalent to
\be
   v_m = \sum_{i=0}^m v_i \, v_{m-i} = \{J,v_m\} + \sum_{i=1}^{m-1} v_i \, v_{m-i}
   \qquad\quad  m=0,1,2, \ldots  \; .  \label{vm_V2=I}
\ee
We further assume that the first of the hierarchy equations (\ref{gAKNS}), i.e.
$V_x = [\la J +v_1,V]$ holds, and thus (\ref{gAKNS1}).
Together with (\ref{vm_V2=I}), this implies
\be
   v_{m+1} = - ( v_{m,x} + \sum_{i=1}^m v_i \, v_{m+1-i} - [v_1,v_m] ) \, H
\ee
where $H := 2J-I$ which satisfies $H^2=I$. This allows us to express $v_m$, $m>1$,
iteratively in terms of $u := v_1$ and its derivatives with respect to $x$,
\be
    v_2 &=& -(u_x+u^2) \, H \, , \qquad
    v_3 = u_{xx} - 2 \, u^3 + [u,u_x]   \\
    v_4 &=& -(u_{xxx} + \{u,u_{xx}\} - 3 \, \{ u^2,u_x\}  -(u_x)^2 - 3 \, u^4 ) \, H
\ee
etc. It is well-known that soliton equations emerge from the hierarchy (\ref{gAKNS})
for $n>1$. But now we show how to obtain them from identities in $\cal I$.
Using the above results, the $\Phi$-images (\ref{Phi(PkPn)}) of algebraic
identities for $n=1$ and $k=2,3$ become
\be
     v_{2,y} \, J = v_{3,x} \, J \, , \qquad
     v_{2,t} \, J = v_{4,x} \, J
        \label{red:Phi(P-P)}
\ee
where $y := t_2$ and $t:=t_3$. Let us choose
\be
    J = \left(\begin{array}{cc} 1&0 \\ 0&0 \end{array}\right) \, , \qquad
    u = \left(\begin{array}{cc} 0&q \\ r&0 \end{array}\right)
\ee
where $q$ and $r$ are elements of a (not necessarily commutative) algebra. Then
\be
    v_2 &=& \left(\begin{array}{cc} -q r & q_x \\ -r_x & r q \end{array}\right) \, , \qquad
    v_3 = \left(\begin{array}{cc} q r_x - q_x r & q_{xx} - 2 \, q r q \\
                                  r_{xx} - 2 \, r q r & r q_x - r_x q \end{array}\right) \\
    v_4 &=& \left(\begin{array}{cc}
        -q r_{xx} - q_{xx} r + q_x r_x +3 q r q r & -q_{xxx} + 3 \, (q_x r q + q r q_x) \\
        -r_{xxx} + 3 \, (r_x q r + r q r_x) & r q_{xx} + r_{xx} q - r_x q_x - 3 \, r q r q
     \end{array}\right)
\ee
and equations (\ref{red:Phi(P-P)}) yield (after an integration)
\be
    (q_y - q_{xx} + 2 \, q r q) \, r = 0 \, ,&& \quad
    r_y + r_{xx} - 2 \, r q r = 0              \label{AKNS1a}  \\
    (q_t - q_{xxx} + 3 \, (q_x r q + q r q_x) ) \, r = 0 \, ,&& \quad
    r_t - r_{xxx} + 3 \, (r_x q r + r q r_x) = 0  \; . \qquad   \label{AKNS1b}
\ee
(\ref{AKNS1a}) is a system of coupled nonlinear Schr\"odinger equations.
(\ref{AKNS1b}) yields with $r=1$ the (noncommutative) KdV equation,
and with $q=r$ the (noncommutative) mKdV equation. Moreover,
(\ref{KP-id-gAKNS}) is satisfied as a consequence of (\ref{AKNS1a}) and (\ref{AKNS1b}).

\subsection{$V^3=I$ reduction}
In this subsection we sketch another reduction: $V^3=I$. Let
\be
    J = \left(\begin{array}{ccc} 1&0&0 \\ 0&\zeta&0\\0&0&\zeta^2 \end{array}\right) \, ,
    \qquad
    u := v_1 = (1 + 2 \, \zeta) \left(\begin{array}{ccc} 0&r&q \\
               q&0&r\\r&q&0 \end{array}\right)
\ee
where $\zeta$ is a third root of unity (so that $\zeta^2+\zeta+1=0$) and $q,r \in \B$.
(\ref{gAKNS1}) determines the non-diagonal part of $v_m$, the reduction the
diagonal part. We obtain
\be
   v_2 &=& \left(\begin{array}{ccc} 3 \, q \,r & \zeta \, r_x & \zeta/(1+\zeta) \, q_x \\
         - \zeta \, q_x & - 3 \, (1+\zeta) \, q \,r & r_x \\
         -\zeta/(1+\zeta) \, r_x & -q_x & - 3/(1+\zeta) \, q \,r \end{array}\right)  \\
   v_3 &=& \left(\begin{array}{ccc}
           (\zeta-1)\, D/(1+\zeta)  & -R/(2+\zeta) & \zeta Q/(1+2\zeta) \\
          - Q/(2+\zeta) & -(1+2 \zeta)\, D/(1+\zeta) & (1+\zeta) R/(\zeta-1) \\
          \zeta R/(1+2\zeta) & (1+\zeta) Q/(\zeta-1) & (1-\zeta) \, D
          \end{array}\right) . \qquad
\ee
where $Q:=q_{xx}+9q^2r-3rr_x$, $R:=r_{xx}+9qr^2+3qq_x$ and $D := q^3 +r^3 + q r_x - q_x r$.
Next we compute (\ref{Phi(P2)}) and evaluate the identity (\ref{Phi(PkPn)}) for $n=1$ and $k=2$.
Setting $t := (1+2 \zeta) \, t_2$, this yields the following system of coupled Burgers equations
\be
    q_t - q_{xx} + 6 \, r \, r_x = 0 \, , \qquad
    r_t + r_{xx} + 6 \, q \, q_x = 0 \; .
\ee

\section{From AKNS to KP}
\label{section:AKNS->KP}
The existence of maps $\Phi_{\mathrm{KP}}$ and $\Phi_{\mathrm{AKNS}}$ which map
identities in the algebra $\A(P)$ to equations of the KP, respectively AKNS hierarchy
suggests a relation between the latter hierarchies. Let us see what happens if we
identify their images.
The equation $\Phi_{\mathrm{KP}}(P) = \Phi_{\mathrm{AKNS}}(P)$ reads
\be
     \phi_x = v_2 \, J    \label{AKNS->KP-constraint}
\ee
and, more generally, $\Phi_{\mathrm{KP}}(P^{\bullet n}) = \Phi_{\mathrm{AKNS}}(P^{\bullet n})$
means
\be
    \phi_{t_n} = (\lambda^n V^n)_{-1} \, J  \; .
\ee
With the reduction treated in section~4.1, (\ref{AKNS->KP-constraint}) becomes
$\phi_x = - q \, r$, which indeed is a well-known (symmetry) constraint of the KP
equation \cite{sconstr}.
As a consequence of it, if $q,r$ satisfy the AKNS equations (\ref{AKNS1a}) and
(\ref{AKNS1b}), then $\phi$ satisfies the potential KP equation (\ref{pKP}).
(\ref{AKNS->KP-constraint}) generalizes this relation to matrix (potential)
KP equations. A thorough analysis of $\Phi_{\mathrm{KP}} = \Phi_{\mathrm{AKNS}}$
has still to be carried out, but we verified with the help of computer algebra in
several examples that indeed matrix (potential) KP equations
are satisfied by (\ref{AKNS->KP-constraint}) as a consequence of the corresponding
AKNS equations. We expect that this relation extends to the whole hierarchies.
We plan to report on the relations between the abstract algebra $\A(P)$ and the
KP and AKNS hierarchies sketched in this work in more detail in a separate publication.

\end {document}